\documentstyle[espcrc2,psfig,fleqn]{article}

\newcommand{\figwidth}{7.2cm}

\begin{document}
\title{
  Perturbative renormalization of the $\Delta B=2$
  four-quark operators in lattice 
  NRQCD\thanks{poster presented by K-I.~Ishikawa.}
  }

\author{
  S.~Hashimoto,\address{
    High Energy Accelerator Research Organization (KEK),
    Tsukuba, Ibaraki 305-0801, Japan}
  K-I.~Ishikawa,$^{\rm a}$
  T.~Onogi,\address{
    Department of Physics, Hiroshima University,
    Higashi-Hiroshima, Hiroshima 739-8526, Japan}
  M.~Sakamoto,$^{\rm b}$
  N.~Tsutsui,$^{\rm a}$ and
  N.~Yamada$^{\rm a}$
  }

\begin{abstract}
  We present a perturbative calculation of the
  renormalization constants for the $\Delta B=2$ four-quark
  operators, which is needed to evaluate the bag parameters
  $B_B$ and $B_S$ in lattice QCD.
  The one-loop coefficients are calculated for the $O(1/M)$
  NRQCD heavy quark action and the $O(a)$-improved Wilson
  action for light quark.
  Using these coefficients, which remove $O(\alpha_s/aM)$
  error, we reanalyze our previous calculation of the bag
  parameters. 
\end{abstract}

\maketitle

\section{Introduction}
In the lattice calculation of the $B$ meson $B$-parameters
\cite{Review}, 
the use of the effective theory for heavy quark is essential
to obtain results with controlled systematic errors.
The static approximation in which the heavy quark mass is
sent to infinity has been used by many authors \cite{HQET},
and the next step is to incorporate the correction from
finite $b$ quark mass.
In this direction, we performed an exploratory lattice
calculation using the NRQCD action for heavy quark, in which
we found that the $1/m_B$ correction could be sizable for
$B_B$ \cite{NRQCD_Hiroshima} and for 
$B_S$ \cite{NRQCD_Hiroshima_BS}.
A drawback in these calculations was, however, that the
perturbative renormalization of the four-quark operators
were not available for the NRQCD action and we used the
one-loop coefficients for the static action.
As a result our results contained large systematic error of
$O(\alpha_s/aM)$, which is about 10\% if evaluated with an
order counting at $\beta$ = 5.9. 

In this paper, we present a perturbative calculations of
renormalization constants of the heavy-light four quark
operators with the $O(1/M)$ lattice NRQCD action.
The results connect the missing link in our previous
calculations, and we present a reanalysis of $B_B$ and $B_S$.
A full description of our results is published in
\cite{This_Work}.

\section{Operator Definitions}

The $B$ parameters in the $B_q-\bar{B}_q$ are defined
through 
\begin{equation}
  \langle \bar{B}_{q}|
  O^{\overline{\rm MS}}_{L}(\mu)
  | B_{q}\rangle
  =  
  \frac{8}{3} M_{B_{q}}^{2} f_{B_{q}}^{2} 
  B^{\overline{\rm MS}}_{L_{q}}(\mu),
\end{equation}
\vspace*{-5mm}
\begin{eqnarray}
  \lefteqn{
    \langle \bar{B}_{q}| 
    O^{\overline{\rm MS}}_{S}(\mu) 
    | B_{q}\rangle
    }
  \nonumber\\
  & = &
  -\frac{5}{3} M_{B_{q}}^{2} f_{B_{q}}^{2}
  B^{\overline{\rm MS}}_{S_{q}}(\mu)
  \left(
    \frac{M_{B_{q}}}{
      \overline{m}_{b}(\mu)+\overline{m}_{q}(\mu)}
  \right)^{2},
  \nonumber\\
\end{eqnarray}
where a subscript $q$ denotes light quark flavor, i.e. $d$
or $s$.
$B_L$ is usually called $B_B$, but we use a notation $B_L$
in this paper to remind that it parameterizes a matrix
element of the operator $O_L$.
In the above equations, the operators 
$O^{\overline{\rm MS}}_{L}(\mu)$ and  
$O^{\overline{\rm MS}}_{S}(\mu)$ are defined in the
continuum theory with the  $\overline{\rm MS}$ scheme.
Their Dirac structure is
\begin{eqnarray}
  O_{L} & = & 
  \bar{b}\gamma_{\mu}P_{L}q \bar{b}\gamma_{\mu}P_{L}q,
  \\
  O_{S} & = & 
  \bar{b}P_{L}q \bar{b}P_{L}q,
\end{eqnarray}
where $P_{L}=1-\gamma_{5}$.
The totally anti-commuting convention for $\gamma_{5}$ is
assumed in the dimensional regularization.

The matching of these continuum operators is done against
the lattice NRQCD.
Namely we consider the $O(1/M)$ NRQCD action for heavy quark
and the $O(a)$-improved Wilson fermion for light quark.
For gluons the standard plaquette action is used.

For $O_{L}$ and $O_{S}$ the following heavy-light four-quark
operators appear in the matching relation:
\begin{eqnarray}
&&\hspace*{-2em}
  O_L =
  \left[ \bar{b}\gamma_\mu P_L q\right]
  \left[ \bar{b}\gamma_\mu P_L q\right],\nonumber\\
&&\hspace*{-2em}
  O_R =
  \left[ \bar{b}\gamma_\mu P_R q\right]
  \left[ \bar{b}\gamma_\mu P_R q\right],\nonumber\\
&&\hspace*{-2em}
  O_S =
  \left[ \bar{b} P_L q\right]
  \left[ \bar{b} P_L q\right],\nonumber\\
&&\hspace*{-2em}
  O_N =
  2 \left[ \bar{b}\gamma_\mu P_L q\right]
    \left[ \bar{b}\gamma_\mu P_R q\right]
  + 4 \left[ \bar{b} P_L q\right]
      \left[ \bar{b} P_R q\right],\nonumber\\
&&\hspace*{-2em}
  O_M =
  2 \left[ \bar{b}\gamma_\mu P_L q\right]
    \left[ \bar{b}\gamma_\mu P_R q\right]
  - 4 \left[ \bar{b} P_L q\right]
      \left[ \bar{b} P_R q\right],\nonumber\\
&&\hspace*{-2em}
  O_P =
  2 \left[ \bar{b}\gamma_\mu P_L q\right]
    \left[ \bar{b}\gamma_\mu P_R q\right]
  +12 \left[ \bar{b} P_L q\right]
      \left[ \bar{b} P_R q\right],\nonumber\\
&&\hspace*{-2em}
  O_T =
  (2+N_c)
  \left[ \bar{b}\gamma_\mu P_L q\right]
  \left[ \bar{b}\gamma_\mu P_R q\right] \nonumber \\
&&\hspace*{2em}
  -2(3 N^2_c - 2 N_c - 4)
  \left[ \bar{b} P_L q\right]
  \left[ \bar{b} P_R q\right],
\end{eqnarray}
where $P_{R} = 1+\gamma_5$.
The heavy quark field $b$ has four Dirac components, and it
is obtained from the two-component spinor non-relativistic
quark (anti-quark), $Q$ ($\chi$), through the
Foldy-Wouthuysen-Tani transformation,
\begin{equation}
  \label{eq:FWT}
  b =
  \left[
    1 - \frac{\vec{\gamma}\cdot\vec{D}}{2M}
  \right]
  \left(
    \begin{array}{c}
      Q \\ \chi^{\dag}
    \end{array}
  \right).
\end{equation}

In the static limit $M\rightarrow\infty$, where algebra is
greatly reduced, we consider the $O(a)$-improvement of the
lattice four-quark operators through $O(\alpha_s a)$.
In doing that we also define the following dimension seven
operators: 
\begin{eqnarray}
  O_{LD} 
  & = &
  \left[ \bar{b}\gamma_\mu P_L q\right]
  \left[ \bar{b}\gamma_\mu P_L
    (a \vec{D}\cdot\vec{\gamma} )q\right],\nonumber\\
  O_{SD} 
  & = &
  \left[ \bar{b} P_L q\right]
  \left[ \bar{b} P_L (a \vec{D}\cdot\vec{\gamma} )q\right],\nonumber\\
  O_{ND}
  & = &
  2 \left[ \bar{b}\gamma_\mu P_L q\right]
    \left[ \bar{b}\gamma_\mu P_R
      (a \vec{D}\cdot\vec{\gamma} )q\right]\nonumber \\
  & &
  + 4 \left[ \bar{b} P_L q\right]
      \left[ \bar{b} P_R
        (a \vec{D}\cdot\vec{\gamma})q\right], \nonumber\\
  O_{PD}
  & = &
  2 \left[ \bar{b}\gamma_\mu P_L q\right]
    \left[ \bar{b}\gamma_\mu P_R
      (a \vec{D}\cdot\vec{\gamma} )q\right]\nonumber \\
  & &
  +12 \left[ \bar{b} P_L q\right]
      \left[ \bar{b} P_R
        (a \vec{D}\cdot\vec{\gamma})q\right].
\end{eqnarray}

\section{Matching}
At one-loop level, the matching of the continuum operator
$O^{\overline{\rm MS}}_{X}(\mu)$ is given by
\begin{eqnarray}
  O^{\overline{\rm MS}}_{X}(\mu)
  & = &
  \left[
    1+\frac{\alpha_s}{4\pi}{ \zeta_{X,X}(\mu;1/a)}
  \right] O^{\rm lat}_{X}(1/a)
  \nonumber \\
  & & 
  + \sum_{Y={\rm dim.} 6}
    \frac{\alpha_s}{4\pi}{\zeta_{X,Y}(\mu;1/a)}
    O^{\rm lat}_{Y}(1/a)
  \nonumber \\
  & &
  + \sum_{Z={\rm dim.} 7}
    \frac{\alpha_s}{4\pi}\zeta_{X,Z}
    O^{\rm lat}_{Z}(1/a),
\end{eqnarray}
where the summations with $Y$ and $Z$ run over dimension six
and seven operators respectively. 
$\mu$ is the renormalization scale in the $\overline{\rm
  MS}$ scheme and $a$ is lattice spacing.

The coefficients $\zeta$ are determined so that the on-shell 
scattering amplitudes in the continuum and lattice theories
agree with each other.
For the external state, zero spatial momentum heavy and
light quark suffice to determine the coefficients for
dimension six operators.
A derivative in terms of spatial momenta have to be
considered to obtain the dimension seven coefficients.

\subsection{$O_{L}$}

\begin{figure}[t]
  \begin{center}
    \leavevmode\psfig{file=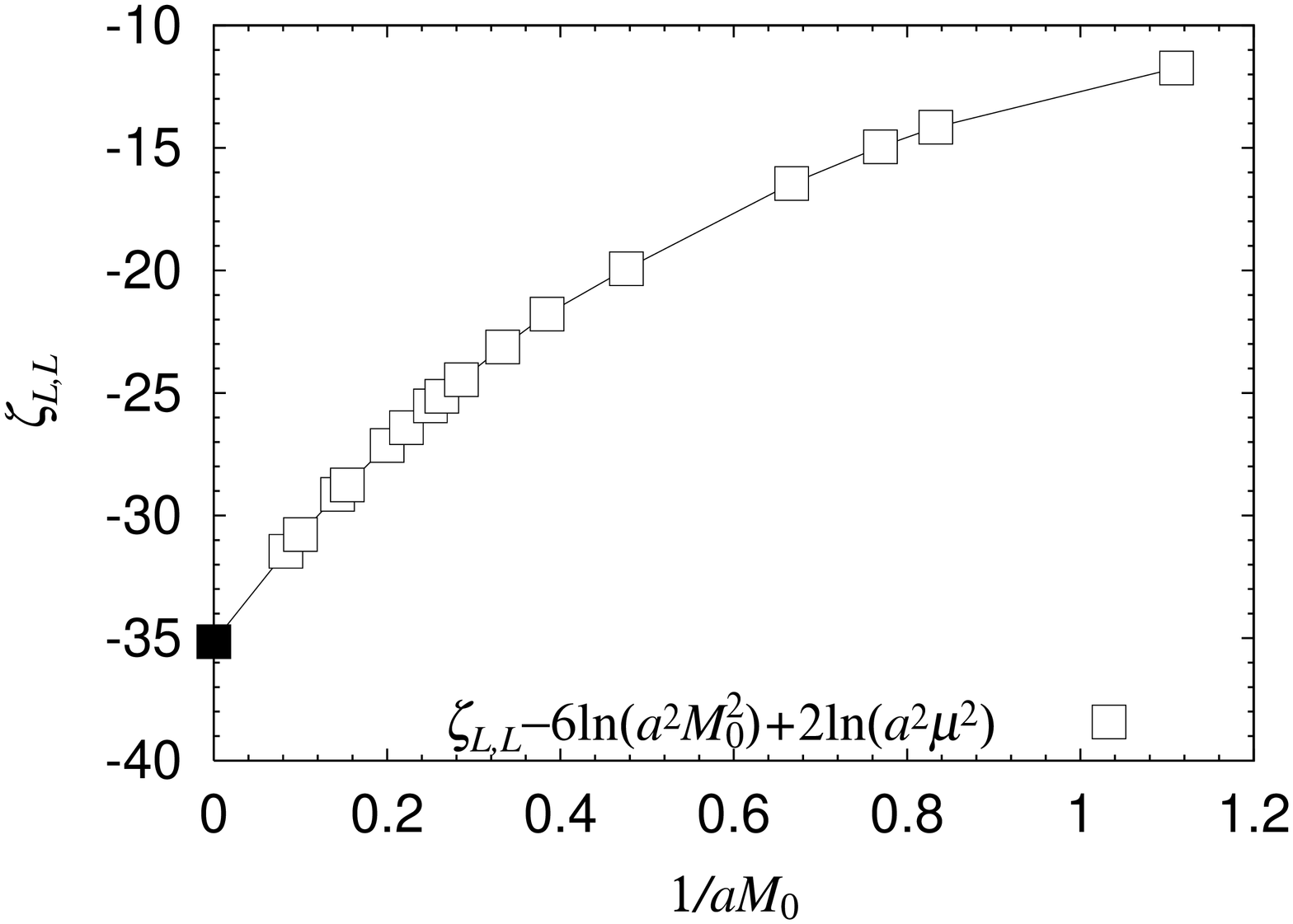,width=\figwidth,angle=-90}
    \leavevmode\psfig{file=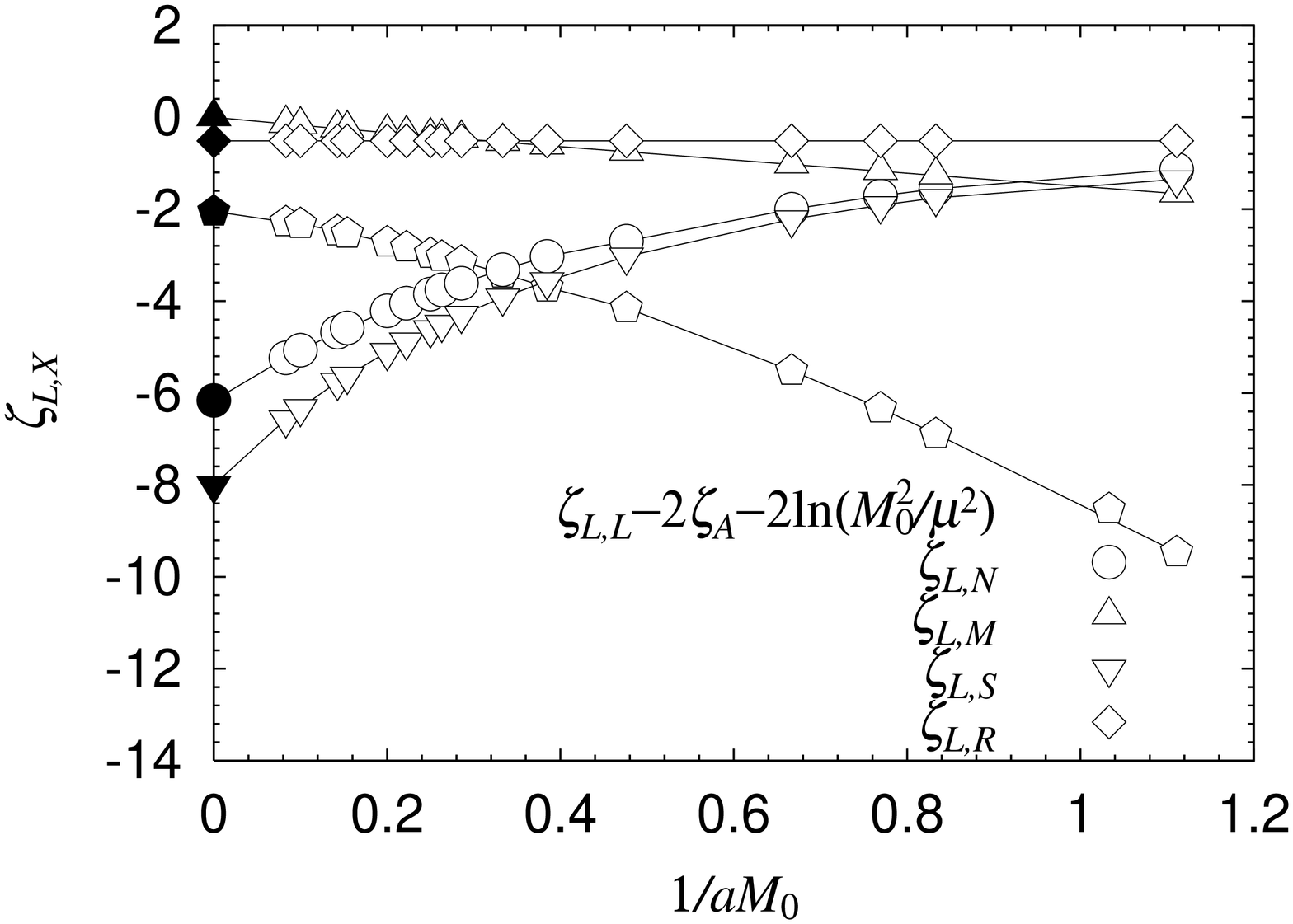,width=\figwidth,angle=-90}
    \vspace*{-2.5em}
    \caption{
      Mass dependence of $\zeta_{L,X}$. 
      $\zeta_{A}$ is the matching coefficient for the
      temporal component of the heavy-light axial vector
      current.
      } 
    \label{fig:zeta_LX}
    \vspace*{-2.0em}
  \end{center}
\end{figure}

The matching relation for $O_{L}$ is
\begin{eqnarray}
  O^{\overline{\rm MS}}_{L}(\mu)
  & = &
  \left[ 1 + \frac{\alpha_s}{4\pi} \zeta_{L,L}(\mu;1/a) \right]
  O_{L}^{\rm lat}(1/a) \nonumber \\
  & &
  + \frac{\alpha_s}{4\pi} \zeta_{L,S}(1/a)\ 
  O_{S}^{\rm lat}(1/a) \nonumber \\
  & &
  + \frac{\alpha_s}{4\pi} \zeta_{L,N}(1/a)\ 
  O_{N}^{\rm lat}(1/a) \nonumber \\
  & &
  + \frac{\alpha_s}{4\pi} \zeta_{L,M}(1/a)\ 
  O_{M}^{\rm lat}(1/a) \nonumber \\
  & &
  + \frac{\alpha_s}{4\pi} \zeta_{L,R}(1/a)\ 
  O_{R}^{\rm lat}(1/a) \nonumber \\
  & &
  + \frac{\alpha_s}{4\pi} \zeta_{L,LD}
  O_{LD}^{\rm lat}(1/a) \nonumber \\
  & &
  + \frac{\alpha_s}{4\pi} \zeta_{L,ND}\
  O_{ND}^{\rm lat}(1/a).
  \label{eqOL}
\end{eqnarray}
The coefficient $\zeta_{L,L}(\mu;1/a)$ of the leading
operator $O_L^{\rm lat}$ contains logarithmic terms
$6\ln(a^2M_0^2)-2\ln(a^2\mu^2)$, while others do not have
$\mu$ dependence.
The operator $O_M$ does not exist in the matching of the
static operator, but it becomes necessary for NRQCD.

Figure~\ref{fig:zeta_LX} shows the heavy quark mass
dependence of $\zeta_{L,X}$ for the dimension six operators. 
Filled symbols at $1/aM_{0}=0$ are calculated for the static
action. 
We find a large slope in $1/aM_0$ for $\zeta_{L,L}$ as shown
in the top panel of Figure~\ref{fig:zeta_LX}.
Most of the mass dependence comes from factorized diagrams,
i.e. the four-quark operator can be split into two
axial-vector currents and the gluon propagator does not 
connect these two. 
As a result, the coefficient $\zeta_{L,L}$ is much reduced
if we subtract the contributions of two axial vector
current, as plotted in the lower panel
($\zeta_{L,L}-2\zeta_A$). 
This combination will appear in the analysis of the $B$
parameters, when we consider the ratio 
$\langle O_L\rangle/\langle A_0\rangle^2$.

The coefficients for the dimension seven operators are
obtained in the static limit as
$\zeta_{L,LD}$ = $-$17.20 and 
$\zeta_{L,ND}$ = $-$9.20
\cite{O_alpha_a_static}.

\subsection{$O_{S}$}

\begin{figure}[t]
  \begin{center}
    \leavevmode\psfig{file=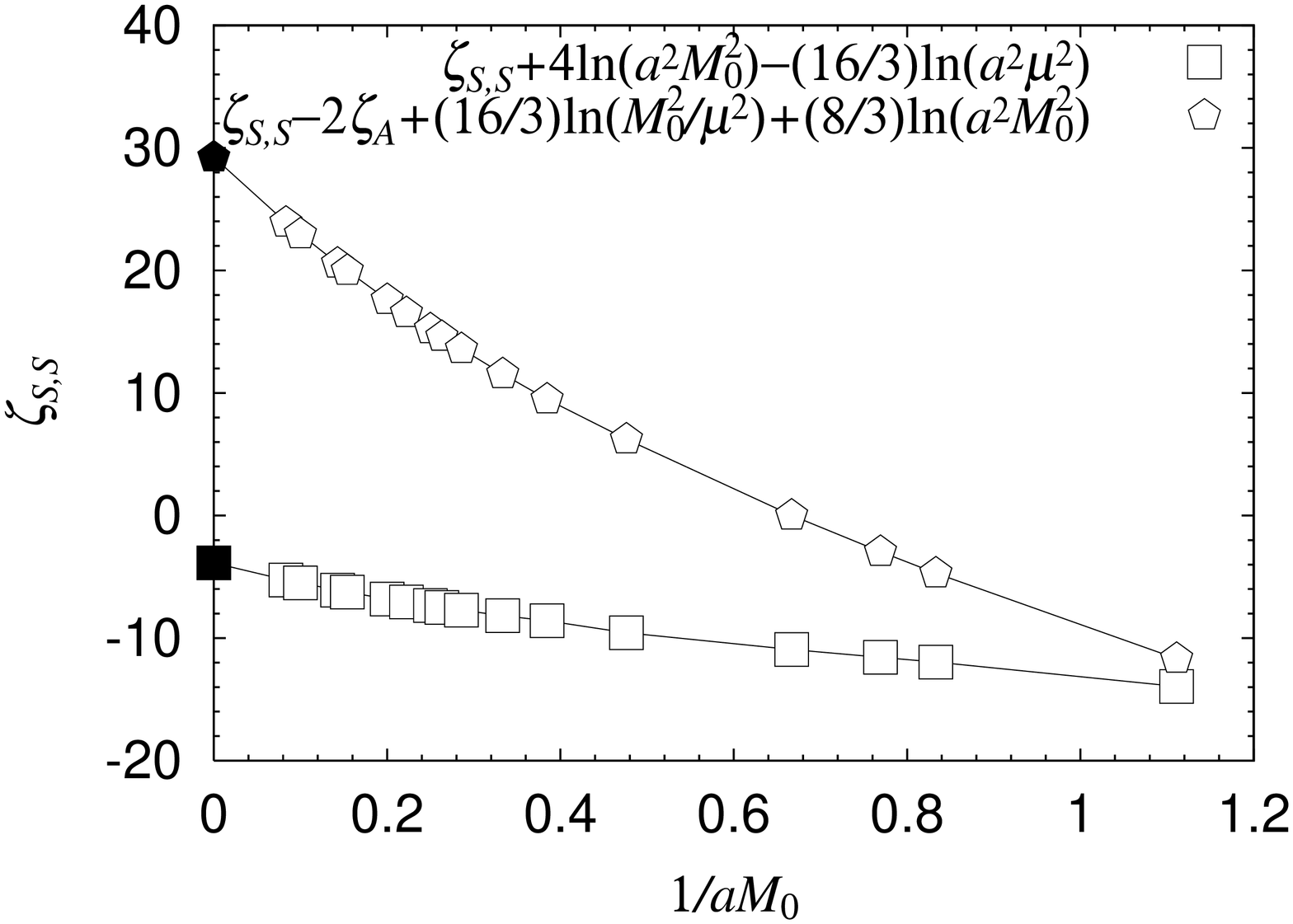,width=\figwidth,angle=-90}
    \leavevmode\psfig{file=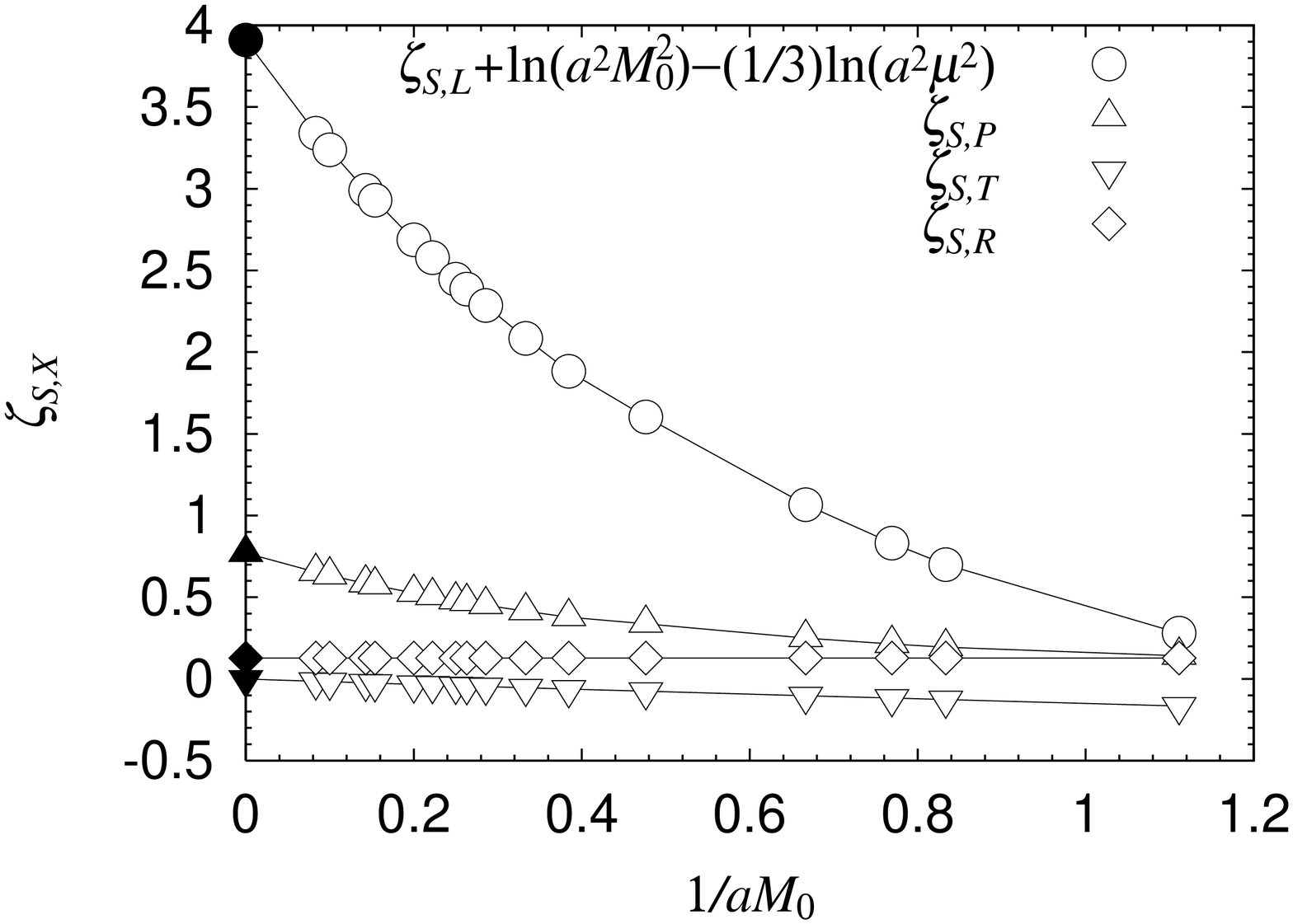,width=\figwidth,angle=-90}
    \vspace*{-2.5em}
    \caption{Mass dependence of $\zeta_{S,X}$.}
    \label{fig:zeta_SX}
    \vspace*{-2.0em}
  \end{center}
\end{figure}

The matching of $O_S$ is given by
\begin{eqnarray}
  O_S^{\overline{\rm MS}}(\mu)
  & = &
  \left[ 1 + \frac{\alpha_s}{4 \pi}\zeta_{S,S}(\mu;1/a) \right]
  O_S^{\rm lat}(1/a)  \nonumber \\
  & &
  + \frac{\alpha_s}{4 \pi} \zeta_{S,L}(\mu;1/a)
    O_L^{\rm lat}(1/a) \nonumber\\
  & &
  + \frac{\alpha_s}{4 \pi} \zeta_{S,P}(1/a)
    O_P^{\rm lat}(1/a) \nonumber \\
  & &
  + \frac{\alpha_s}{4 \pi} \zeta_{S,R}(1/a)
    O_R^{\rm lat}(1/a) \nonumber \\
  & &
  + \frac{\alpha_s}{4 \pi} \zeta_{S,T}(1/a)
    O_T^{\rm lat}(1/a) \nonumber\\
  & &
  + \frac{\alpha_s}{4 \pi} \zeta_{S,SD}
    O_{SD}^{\rm lat}(1/a) \nonumber \\
  & &
  + \frac{\alpha_s}{4 \pi} \zeta_{S,LD}
    O_{LD}^{\rm lat}(1/a) \nonumber \\
  & &
  + \frac{\alpha_s}{4 \pi} \zeta_{S,PD}
    O_{PD}^{\rm lat}(1/a),
  \label{eqOS}
\end{eqnarray}
where $O_T$ is new for the NRQCD action.
The coefficients $\zeta_{S,X}$ are plotted in
Figure~\ref{fig:zeta_SX}. 
Like $\zeta_{L,L}$ in the matching of $O_L$, the leading
operator $O_S$ has large coefficient, which is reduced by
subtracting the factorized contribution, as shown in the
upper panel.

The coefficients for the dimension seven operators are
$\zeta_{S,SD}$ = $-$6.88,
$\zeta_{S,LD}$ = 2.58 and $\zeta_{S,PD}$ = 1.15
in the static limit.

\section{Reanalysis of $B$ Parameters}
With these perturbative coefficients, we reanalyze our
previous lattice calculation for the $B$ parameters, $B_{L}$
and $B_{S}$ \cite{NRQCD_Hiroshima,NRQCD_Hiroshima_BS}.

\subsection{$B_{L}$}
The $B$ parameter $B_{L}$ is calculated through
\begin{equation}
  B^{\overline{\rm MS}}_{L}(\mu)=
  \sum_X
  Z_{L,X/A^{2}}(\mu;1/a)B^{\rm lat}_{X}(1/a),
\end{equation}
where $X$ runs over $L$, $S$, $N$, $M$ and $R$.
The renormalization constants are
\begin{eqnarray}
  Z_{L,L/A^2}(\mu;1/a)
  & = &
  \nonumber \\
  \lefteqn{
    1+\frac{\alpha_s}{4\pi}
    \left(
      \zeta_{L,L}(\mu;1/a)-2\zeta_{A}(1/a)
    \right),
    }
  \nonumber \\
  Z_{L,S/A^2}(1/a)
  & = &
  \frac{\alpha_s}{4\pi}\zeta_{L,S}(1/a), \nonumber \\
  Z_{L,N/A^2}(1/a)
  & = &
  \frac{\alpha_s}{4\pi}\zeta_{L,N}(1/a), \nonumber \\
  Z_{L,M/A^2}(1/a)
  & = & \frac{\alpha_s}{4\pi}\zeta_{L,M}(1/a), \nonumber \\
  Z_{L,R/A^2}(1/a)
  & = & \frac{\alpha_s}{4\pi}\zeta_{L,R}(1/a), \nonumber
\end{eqnarray}
and the lattice $B$ parameters
$B^{\rm lat}_{X_{q}}(1/a)$ were measured in the numerical
simulations through
\begin{eqnarray}
  &&\hspace*{-2em}
  B^{\rm lat}_{X_q}(1/a)=
  \frac{
    \langle \bar{B}_q|O^{\rm lat}_{X}(1/a)|B_q\rangle }
  {\frac{8}{3}
    \langle \bar{B}_q|A^{\rm lat}_4(1/a)|0\rangle
    \langle         0|A^{\rm lat}_4(1/a)|B_q\rangle},
  \nonumber \\
\end{eqnarray}
in our previous paper
\cite{NRQCD_Hiroshima,NRQCD_Hiroshima_BS}.  
Since the dimension seven operators are not included in the
calculation, we could not remove the error of 
$O(\alpha_s a\Lambda_{\rm QCD})$.

\begin{figure}[t]
  \begin{center}
    \leavevmode\psfig{file=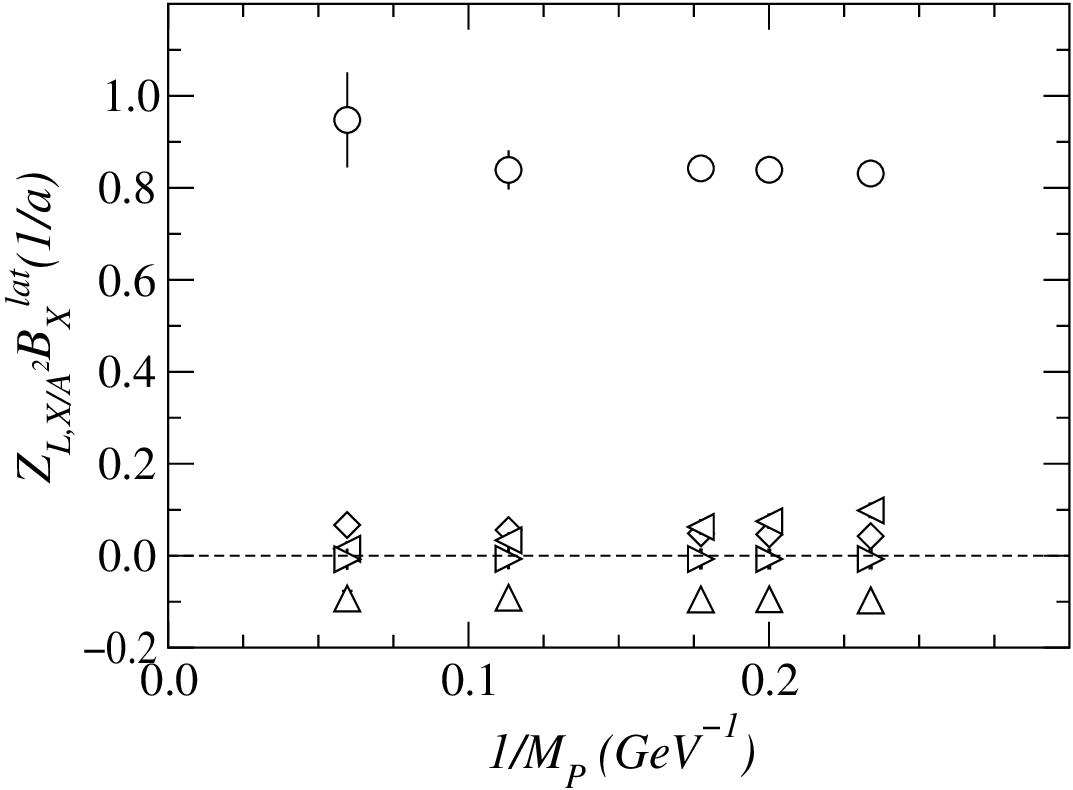,width=\figwidth,clip=}
    \vspace*{-2.5em}
    \caption{Contribution of individual operators
      $Z_{L,X/A^2} B_X^{lat}(1/a)$ to $B_L(\mu)$,
      where $X = L$ ({\Large$\circ$}'s), $S$ ($\Diamond$'s), 
      $R$ ({\Large$\triangleright$}'s), $N$ ($\triangle$'s),
      $M$ ({\Large$\triangleleft$}'s).}
    \label{fig:Z_L,X*B_X}
    \vspace*{-2.0em}
  \end{center}
\end{figure}

Figure~\ref{fig:Z_L,X*B_X} shows the individual
contributions $Z_{L,X/A^{2}}(\mu;1/a)B^{\rm lat}_{X}(1/a)$
for the $B$ parameter $B_{L}$.
The $V$-scheme coupling constant $\alpha_{V}(q^{*})$ is used 
with $q^{*}=2/a$.
Although there is a sizable mass dependence in
$Z_{L,X/A^{2}}(\mu;1/a)$ as we can see from the plot of
coefficients (Figure~\ref{fig:zeta_LX}), it is canceled by
the $1/m_B$ corrections in the matrix elements on the
lattice $B^{\rm lat}_{X}(1/a)$, and there is little mass
dependence in the products shown in
Figure~\ref{fig:Z_L,X*B_X}. 

\begin{figure}[t]
  \begin{center}
    \leavevmode\psfig{file=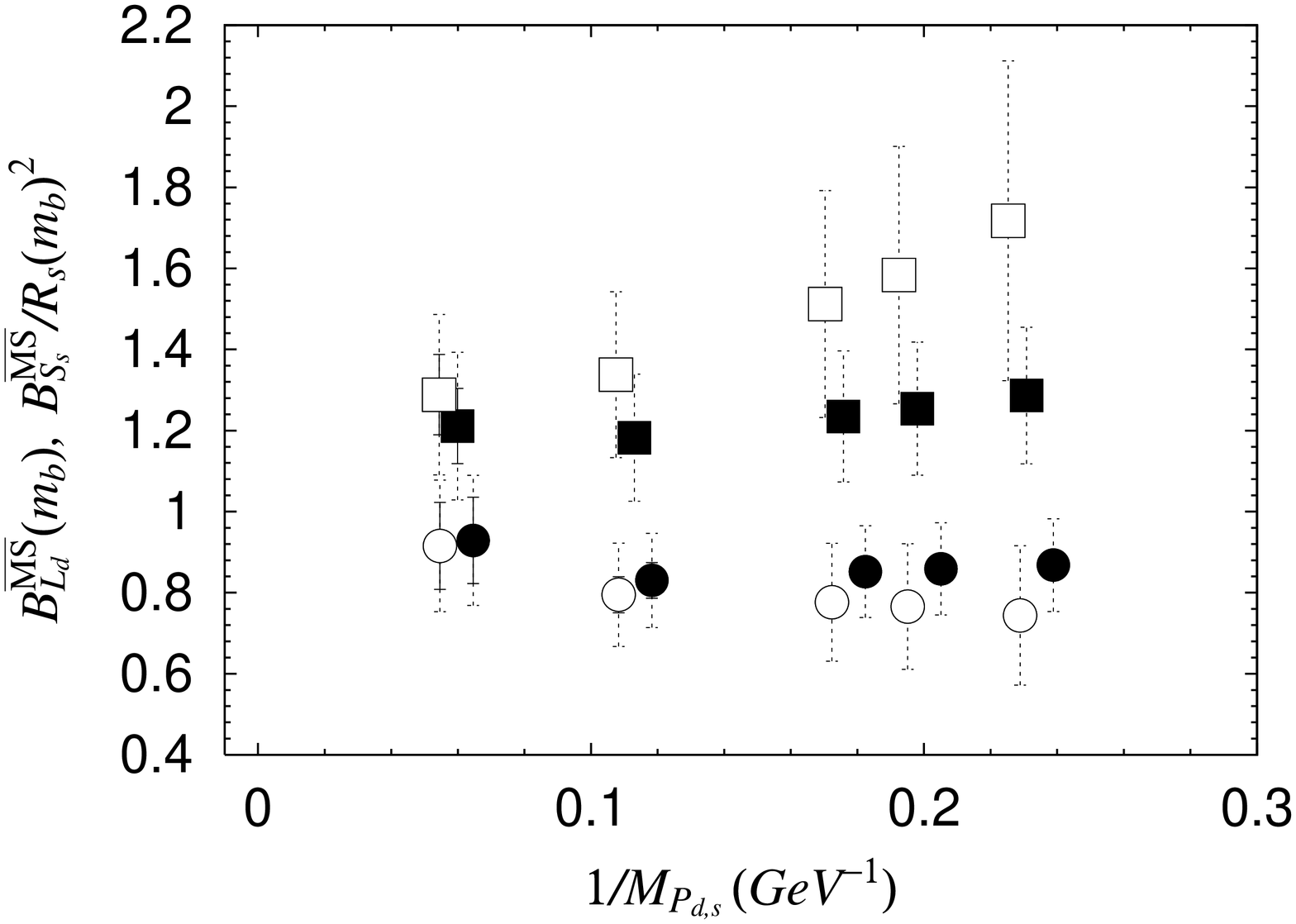,width=\figwidth,angle=-90}
    \vspace*{-2.5em}
    \caption{
      $1/M_P$ dependence of $B^{\overline{\rm MB}}_{L_d}(m_b)$ (circles) and 
      $B^{\overline{\rm MS}}_{S_s}(m_b)/{\cal R}_{s}(m_b)^2$ (squares).
      Solid error bar shows statistical error, dotted includes systematic.}
    \label{fig:B_L}
    \vspace*{-2.0em}
  \end{center}
\end{figure}

Our result for $B^{\overline{\rm MS}}_{L}(\mu)$ is plotted
in Figure~\ref{fig:B_L} as a function of $1/M_P$ with filled
circles. 
The statistical error shown by solid error bars is very
small except for the heaviest heavy quark, while the
systematic error given by dotted error bars is more
significant. 
We estimate the systematic errors with an order counting, in
which the systematic errors of
$O(\alpha_s^2)$,
$O(\alpha_s a\Lambda_{\rm QCD})$,
$O(\alpha_s\Lambda_{\rm QCD}/M)$,
$O((a\Lambda_{\rm QCD})^2)$, and 
$O((\Lambda_{\rm QCD}/M)^2)$
are considered and added in quadrature.
For comparison, we also show a result with one-loop
coefficients calculated for the static heavy quark by open
circles, for which a large systematic error of
$O(\alpha_s/(aM))$ is added.
The effect of the NRQCD renormalization constants is
sizable, but still it is within the estimated systematic
error. 

The numerical results for $B_L$ at physical $B$ meson mass
are 
\begin{eqnarray}
  B^{\overline{\rm MS}}_{L_{d}}(m_{b}) & = & 0.85(3)(11),\\
  B^{\overline{\rm MS}}_{L_{s}}(m_{b}) & = & 0.87(2)(11),
\end{eqnarray}
where the first error is statistical and the second is
systematic one ($\sim 13\%$).
The SU(3) breaking effect is 
\begin{equation}
  \frac{B^{\overline{\rm MS}}_{L_{s}}(m_{b})}
       {B^{\overline{\rm MS}}_{L_{d}}(m_{b})}=1.01(1),
\end{equation}
where only the statistical error is quoted.

\subsection{$B_{S}$}

Using Eq.~(\ref{eqOS}), $B_{S}$ is obtained as
\begin{equation}
  \frac{B^{\overline{\rm MS}}_{S_q}(\mu)}{{\cal R}_{q}(\mu)^2}
  =
  \sum_X Z_{S,X/A^2}(\mu;1/a)B^{'{\rm lat}}_{X_q}(1/a),
  \label{eq:BSMStolat}
\end{equation}
where $X$ runs over $S$, $L$, $R$, $P$ and $T$.
In the LHS, we define a ratio of heavy-light matrix elements
\begin{equation}
  \label{eq:R}
  {\cal R}_{q}(\mu)=
  \left|
    \frac{\langle 0|A^{\overline{\rm MS}}_{4} |B_{q}\rangle}{
          \langle 0|P^{\overline{\rm MS}}(\mu)|B_{q}\rangle}
  \right|.
\end{equation}
The renormalization constants are
\begin{eqnarray}
  Z_{S,S/A^2}(\mu;1/a)
  & = & \nonumber \\
  \lefteqn{
    1+\frac{\alpha_s}{4\pi}
    \left(
      \zeta_{S,S}(\mu;1/a)
      -2\zeta_{A}(1/a)
    \right),
    }
  \nonumber \\
  Z_{S,L/A^2}(\mu;1/a)
  & = &
  \frac{\alpha_s}{4\pi}\zeta_{S,L}(1/a),
  \nonumber \\
  Z_{S,R/A^2}(1/a)
  & = &
  \frac{\alpha_s}{4\pi}\zeta_{S,R}(1/a), \nonumber \\
  Z_{S,P/A^2}(1/a)
  & = &
  \frac{\alpha_s}{4\pi}\zeta_{S,P}(1/a), \nonumber \\
  Z_{S,T/A^2}(1/a)
  & = &
  \frac{\alpha_s}{4\pi}\zeta_{S,T}(1/a). 
  \nonumber
\end{eqnarray}
For $B_S$ we define the lattice $B$ parameters as 
\begin{eqnarray}
  \lefteqn{B^{'{\rm lat}}_{X_q}(1/a) = }
  \nonumber \\
  & &
  \frac{
    \langle \bar{B}_q|O^{\rm lat}_{X}(1/a)|B_q\rangle }
  {-\frac{5}{3}
    \langle \bar{B}_q|A^{\rm lat}_4(1/a)|0  \rangle
    \langle         0|A^{\rm lat}_4(1/a)|B_q\rangle}.
\end{eqnarray}

The mass dependence of 
$B^{\overline{\rm MS}}_{S_s}/{\cal R}_{s}$ is plotted 
in Figure~\ref{fig:B_L} (filled squares).
Open squares are our previous results calculated with the
static renormalization constants.
The mass dependence is less significant in the results with
the correct (NRQCD) renormalization constants.

The numerical results are
\begin{eqnarray}
  \frac{B^{\overline{\rm MS}}_{S_d}(m_{b})}{{\cal R}_{d}(m_{b})^{2}}
  & = & 1.24(3)(16), \\
  \frac{B^{\overline{\rm MS}}_{S_s}(m_{b})}{{\cal R}_{s}(m_{b})^{2}}
  & = & 1.24(3)(16),
\end{eqnarray}
and
\begin{equation}
  \frac{B^{\overline{\rm MS}}_{S_s}(m_{b}){\cal R}_{d}(m_{b})^{2}}
       {B^{\overline{\rm MS}}_{S_d}(m_{b}){\cal R}_{s}(m_{b})^{2}}
  = 1.003(4).
\end{equation}
for the SU(3) breaking ratio.

\section*{Acknowledgments}
S.H. and T.O. are supported in part by 
the Grants-in-Aid of the Ministry of Education
(Nos. 11740162, 12640279). 
K-I.I. and N.Y. are supported by the JSPS Research Fellowship.



\begin{thebibliography}{99}
\bibitem{Review}
  For recent reviews see, S.~Hashimoto,
  Nucl. Phys. {\bf B} (Proc. Suppl.) {\bf 83-84} (2000) 3;
  C.~Bernard, these proceedings.

\bibitem{HQET}
  A.~K.~Ewing {\it et al.}, Phys. Rev. D {\bf 54} (1996) 3526;
  V.~Gim\'{e}nez and G.~Martinelli, Phys. Lett. {\bf B398} (1997) 135;
  J.~Christensen, T.~Draper, and C.~McNeile, Phys. Rev. D {\bf 56} (1997) 6993;
  V.~Gim\'{e}nez and J.~Reyes, Nucl. Phys. {\bf B545} (1999) 576,
  hep-lat/0009007, these proceedings.

\bibitem{NRQCD_Hiroshima}
  S.~Hashimoto, K-I.~Ishikawa, H.~Matsufuru, T.~Onogi, and N.~Yamada, 
  Phys. Rev. D {\bf 60} (1999) 094503.

\bibitem{NRQCD_Hiroshima_BS}
  S.~Hashimoto, K-I.~Ishikawa, T.~Onogi, and N.~Yamada,
  Phys. Rev. D {\bf 62} (2000) 034504.

\bibitem{This_Work}
  S.~Hashimoto, K-I.~Ishikawa, T.~Onogi, M.~Sakamoto, N.~Tsutsui, and N.~Yamada,
  Phys. Rev. D {\bf 62} (2000) 114502.

\bibitem{O_alpha_a_static}
  K-I.~Ishikawa, T.~Onogi, and N.~Yamada,
  Phys. Rev. D {\bf 60} (1999) 034501,
  Nucl. Phys. {\bf B} (Proc. Suppl.) {\bf 83-84} (2000) 301.

\end{thebibliography}
\end{document}